\documentclass{appolb}
 \usepackage{epsfig}
\begin{document}
\eqsec

\title{QCD Critical Point: Synergy of Lattice \& Experiments\thanks
{Plenary talk at the Strangeness in Quark Matter, 2011, Krakow, Poland.} }

\author{Rajiv V.\ Gavai 
\address{Department of Theoretical Physics, Tata Institute of Fundamental
         Research,\\ Homi Bhabha Road, Mumbai 400005, India.}
}

\maketitle

\begin{abstract}

The freeze-out curve, which describes a vast amount of precise experimental 
data in heavy ion collisions, provides a relation between the colliding
energy and the thermodynamical parameters of the fireball.  The variance, skew 
and kurtosis of the event distribution of baryon number are studied at several
energies of interest through Pad\'e resummations of our Lattice QCD results.  
A smooth behaviour is predicted for three ratios of
these quantities at current RHIC and future LHC energies. Any deviations
from these at the RHIC energy scan would signal the presence of a nearby
critical point. Our lattice results on the critical point do show such a
behaviour.

\end{abstract}


\section{Introduction}
\label{int}

Critical points in a phase diagram in the temperature-density plane are special
for many reasons. Universality of critical indices, diverging correlation
length are some of them. For common substances, such as water or carbon
dioxide, the existence of critical point has been established experimentally,
with its location known rather precisely.  For many of these, however, getting
a theoretical, especially first principles based, computation of their
locations turned out to be illusive still.  I wish to describe to you how
things are different for strongly interacting matter, which is naturally
described by Quantum Chromo Dynamics (QCD) and has an inherently higher energy
scale.  Whether the QCD phase diagram has a critical point in its $T$-$\mu_B$
plane, where $\mu_B$ is the baryonic chemical potential, is therefore a tougher
question to address experimentally.  Thanks to the impressive developments on
the experimental fronts, the Relativistic Heavy Ion Collider (RHIC) at BNL, New
York, and the upcoming facilities such as FAIR at GSI, Darmstadt and NICA at
Dubna, Russia can potentially search for it, as these proceedings reveal.

A variety of models have been successfully tested for hadronic interactions.
These were was also the first set of tools used for getting a glimpse of the
QCD phase diagram.  For instance, using an effective chiral Nambu-Jana Lasinio
type model, a phase diagram was obtained, which suggests \cite{WilRaj} a
critical point to exist in a world with two light quarks and one heavier quark.
One would clearly like to have an {\em ab initio} theoretical evidence for it.
This turns out to be difficult as one usually has to deal with large coupling
constants in the world of (low energy) hadronic interactions.  Non-perturbative
Lattice QCD, defined on a discrete space-time lattice, has proved itself to be
the most reliable technique for extracting such information from QCD.  The
hadron spectrum has been computed and predictions of weak decay constants of
heavy mesons have been made.  Application of this approach to finite
temperature QCD has yielded a slew of thermodynamics determinations, such as
the pressure as a function of temperature.  It is therefore natural to ask
whether lattice QCD can help us in locating the QCD critical point.  The
prospect of finding the critical point experimentally makes it exciting both as
a check of theoretical predictions and as a competition for getting there
first.  Clearly, in view of the complexity of this task, one could turn to
lattice QCD again to see if it can provide any hints for the experimental
search program.  In this talk, I summarise the results of our group in
TIFR, Mumbai in these directions.

Due to the well-known fermion doubling problem, one has to make a compromise in
choosing the quark type for any computation.  Mostly staggered quarks are used
in the lattice simulations at finite temperature and density.  These have an
exact chiral symmetry which provides an order parameter for the entire
$T$-$\mu_B$ plane but unfortunately the flavour and spin symmetry is broken for
them on lattice. A representation of precisely 2 (or 2 +1) quark flavours may
thus be problematic, more so on the coarse lattices one is constrained to
employ.  The existence of the critical point, on the other hand, is expected to
depend crucially on the number of flavours.  Although computationally much more
expensive, domain wall or overlap fermions are better in this regard, as they
do have the correct symmetries for any lattice spacing at zero temperature and
density .  Introduction of chemical potential, $\mu$, for these, however, is
not straight-forward due to their non-locality.  Bloch and Wettig \cite{BlWe}
proposed a way to do this.  Unfortunately, it turns out \cite{BGS} that their
prescription breaks chiral symmetry .  Furthermore, the chiral anomaly for it
depends on $\mu$ unlike in continuum QCD \cite{GS10}.  While both these
deficiencies go away in the continuum limit, the practical problem of the lack
of an order parameter on finite lattices remains.    What is
therefore desperately needed is a formalism with continuum-like (flavour and 
spin) symmetries for quarks at nonzero $\mu$  and $T$ with a well-defined order
parameter on lattice.

Finite density simulations needed for locating a critical point suffer from
another well known problem.  This one is inherited from the continuum theory
itself: the fermion sign problem.  It is a major stumbling block in extending 
the lattice techniques to the entire $T$-$\mu_B$ plane.  Several approaches have
been proposed in the past decade to deal with it.  Let me provide a partial
list:  1) Two parameter re-weighting \cite{FoKa}, 2) Imaginary chemical 
potential \cite{ImMu}, 3) Taylor expansion \cite{TaEx}, 4) Canonical ensemble 
method \cite{CaEn}, and 5) Complex Langevin approach \cite{ComLa}.
We employ the Taylor expansion approach, developed independently by the 
TIFR group \cite{TaEx} and the Bielefeld group \cite{TaEx}, to obtain the 
results discussed in the next section.

\section{Lattice Results}
\label{olr}

Our results were obtained by simulating full QCD with two flavours of staggered
fermions of mass $m/T_c =0.1$ on $N_t \times N_s^3$ lattices, with $N_t=4$ and
$N_s=$ 8, 10, 12, 16, 24 \cite{our1} and a finer $N_t= 6 $ with $N_s=$ 12, 18,
24 \cite{our2}. Earlier work by the MILC collaboration for the smaller $N_t$
lattice had determined $m_\rho/T_c = 5.4 \pm 0.2$ and $m_\pi/m_\rho = 0.31 \pm
0.01$, leading to a Goldstone pion of 230 MeV in our case.  For the finer
lattice, we determined $\beta_c$ as well.  We covered a temperature range of 
0.89 $ \le T/T_c \le  $  1.92  by suitably choosing the range of couplings 
on both lattices.  Our measurements were made typically on 50-200 independent
configurations, separated by the respective autocorrelation lengths.  

From the canonical definitions of number densities $n_i$ and 
susceptibilities   $\chi_{n_u,n_d}$, the QCD pressure $P$ can be seen to have
the following expansion in $\mu$:
\begin{equation}
   \frac{\Delta P}{T^4} \equiv \frac{P(\mu, T)}{T^4} - \frac{P(0, T)}{T^4}
   = \sum_{n_u,n_d} \chi_{n_u,n_d}\;
        \frac{1}{n_u!}\, \left( \frac{\mu_u}{T} \right)^{n_u}\, 
        \frac{1}{n_d!}\, \left( \frac{\mu_d}{T} \right)^{n_d}\, 
\end{equation} 
where the indices $n_u$ and $n_d$ denote the number of derivatives of the 
partition function with respect to the corresponding chemical potentials.
We construct the series for baryonic susceptibility from this expansion.
and look for its radius of convergence as the estimate of the nearest 
critical point. 

Successive estimates for the radius of convergence were obtained by
using the ratio method [$r_{n+1,n+3} = \sqrt{{n(n+1)\chi^{(n+1)}_B}/
{\chi^{(n+3)}_B T^2}}$ ] and the n$^{th}$ root method [$ r_n = 
\left(n!{\chi_B^{(2)}}/{\chi_B^{(n+2)} T^n} \right)^{1/n}$].  We used
terms up to 8th order in $\mu$ for doing so.  A key point to note is
that all coefficients of the series must be positive for the critical point to
be at real $\mu$, and thus physical. We thus first look for this condition to
be satisfied and then look for agreement between the two definitions
above as well as their $n$-independence to locate the critical point. The
detailed expressions for all the terms can be found in \cite{our1} where the
method to evaluate them is also explained.  We use stochastic estimators. For
terms up to the 8$^{th}$ order one needs 20 inversions of $(D +m)$ on $\sim$
500 vectors for a single measurement on a given gauge configuration.  This
makes the computation expensive.  Nevertheless, extension to 10th \& even 12th
order may be possible, especially using new emerging ideas \cite{GS10} which
may save up to 60 \% computer time in the measurements. Adding the chemical
potential simply as $\mu N$, as proposed in\cite{GS10}, one has the advantage
that most derivatives of the quark matrix with $\mu$ are zero except the first.
This reduces the number of terms in each nonlinear susceptibility (NLS) 
appearing in the equation above a lot.  One can show that it still leads to 
essentially the same results\cite{gs1}.

\begin{figure}[htb]
\includegraphics[scale=0.48]{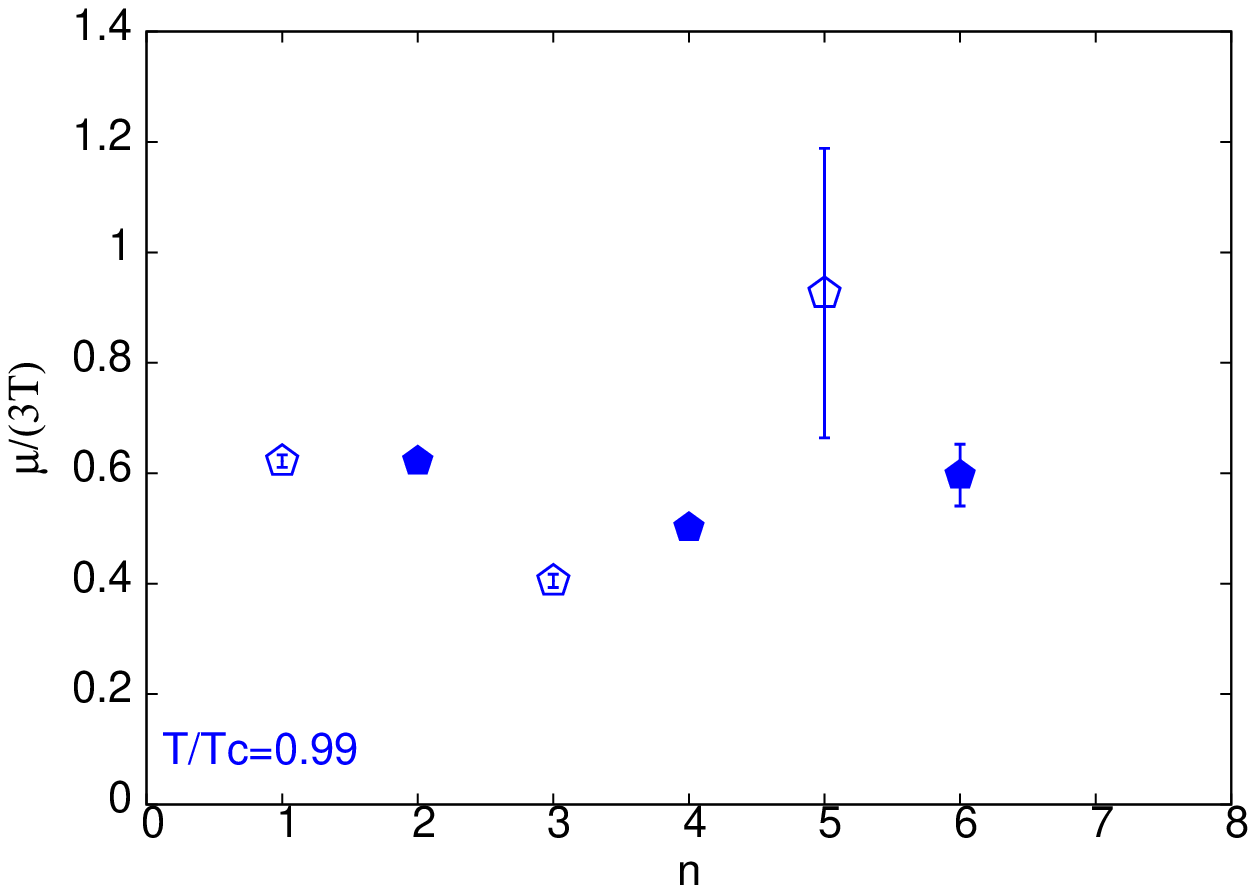}
\includegraphics[scale=0.48]{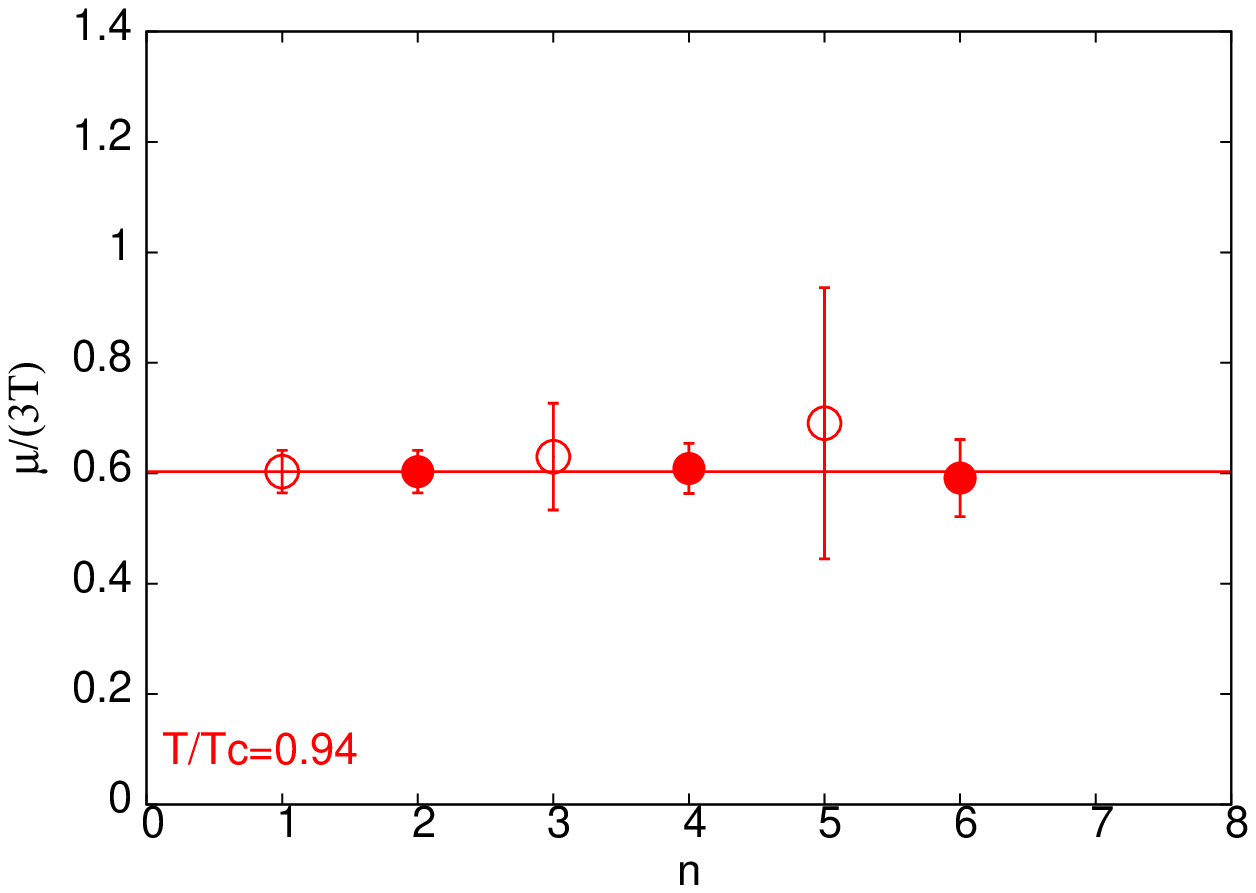}
\caption{Estimates of radii of convergence as a function of the order $n$ and
for our two methods at $T/Tc$ = 0.99 and the critical point $T^E/T_c$ =0.94. 
The open (filled) points display the results for the ratio (root) method.
From Ref. \cite{our2}.  }
\label{rads}
\end{figure}

Fig. \ref{rads} shows our results for both the ratios defined above
on $N_t=6$ lattice at two different temperatures, $T/T_c =0.99$ and 0.94.
All the susceptibilities are positive at both but the ratios fluctuate
for the former while seeming to be constant for the latter.  We\cite{our2}
thus found the coordinates of the endpoint (E)---the critical point---to be
$ T^E/T_c = 0.94\pm0.01$, and $\mu_B^E/T^E = 1.8\pm0.1$
for  the finer lattice. Our earlier coarser \cite{our1} lattice result was
$\mu_B^E/T^E = 1.3\pm0.3$ for similar volumes with an infinite volume result
leading to $\mu_B^E/T^E = 1.1 \pm 0.1$.  In view of this, it may be safer to
quote the critical point to be at $\mu_B/T \sim 1-2$. 

\begin{figure}[htb]
\includegraphics[scale=0.48]{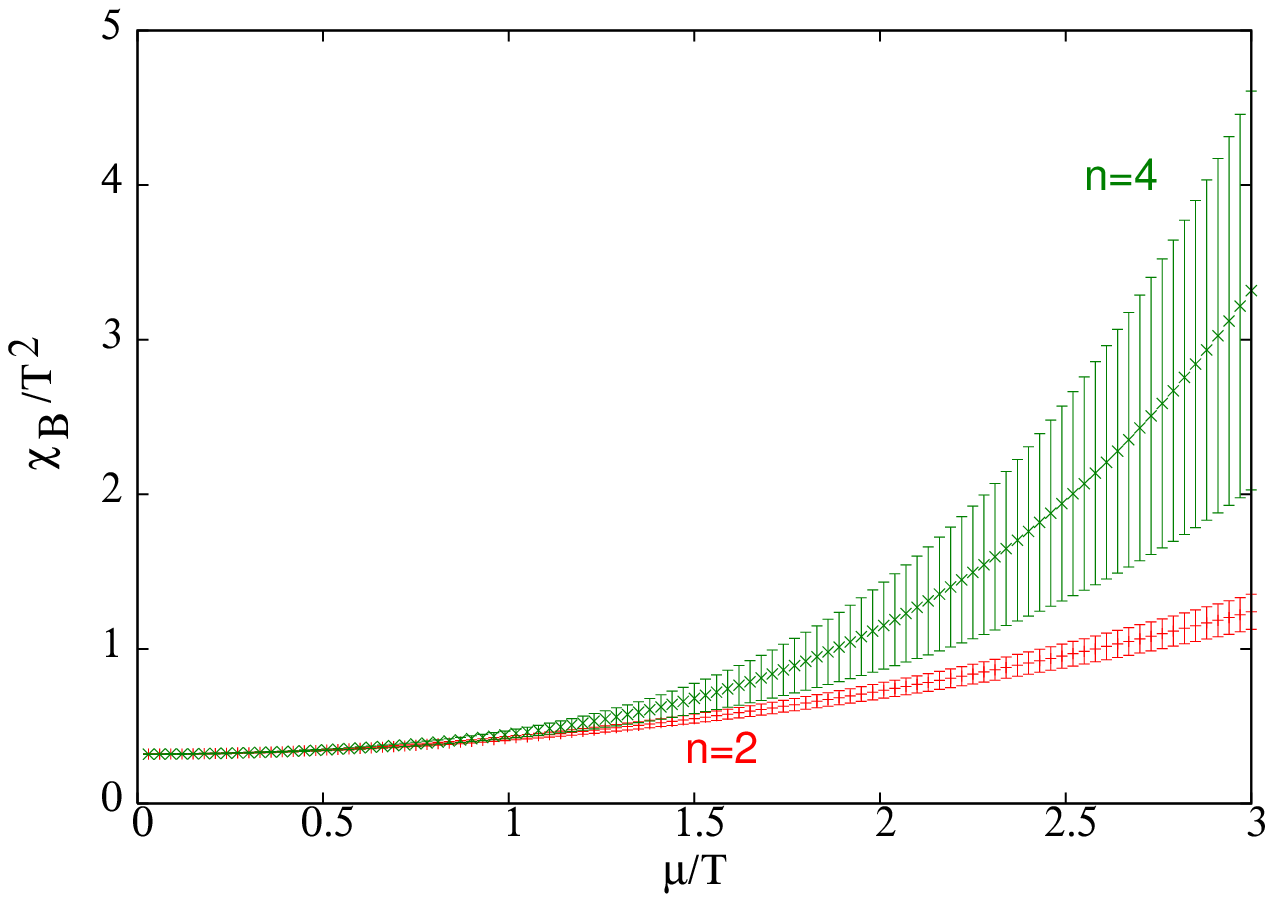}
\includegraphics[scale=0.48]{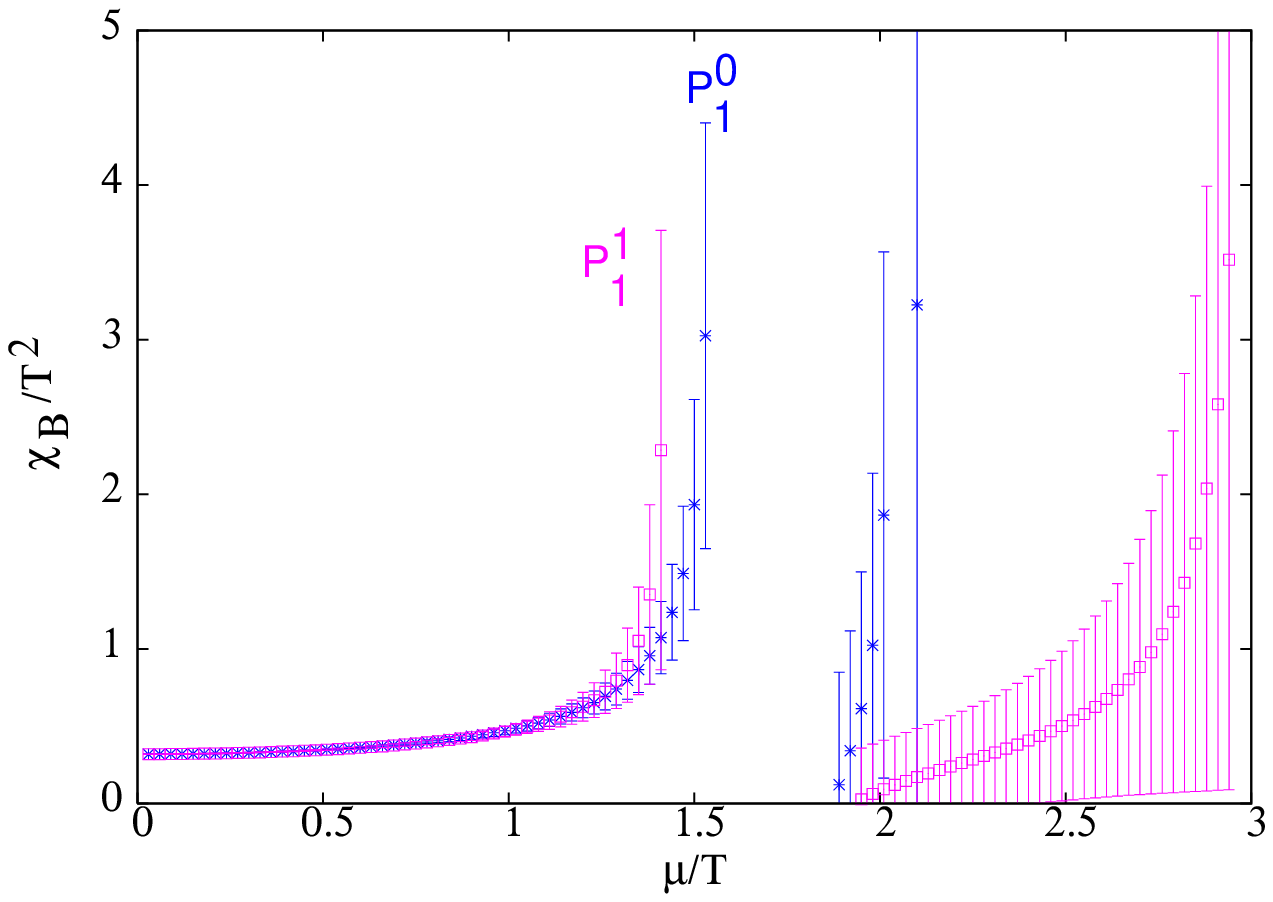}
\caption{The baryonic susceptibility as a function of the baryonic chemical
potential using simple polynomial with all computed orders (left) and Pad\'e
approximants (right).  From Ref. \cite{our2}. }
\label{polypad}
\end{figure}

As a cross check on the location of the $\mu^E/T^E$, one can use the  
series with the numerically determined coefficients directly to construct 
$\chi_B$ for nonzero $\mu$.  As one sees in Fig. \ref{polypad}, it leads to 
smooth curves with no signs of criticality, whereas employing the 
Pad\'e approximants for the series to estimate the radius of convergence
does lead to a consistent window with our above estimate. 

\section{Searching Experimentally}
\label{se}

In order to make predictions for the heavy ion experiments, and design further
any search criteria for the critical point, one needs to know the temperature
$T$ and the chemical potential $\mu_B$ as a function of the colliding energy.
Recently, we proposed \cite{our3} to use the freeze-out curve as a a way to
exploit the information hidden in the nonlinear susceptibilities discussed in
the previous section by evaluating lattice QCD predictions along it.  Recall
the well known observation that the huge amounts of precise data on hadron
abundances in a variety of relativistic heavy ion experiments are well
described \cite{stm}  by statistical thermodynamical models. Indeed this
description has been found to work over the entire range of experiments made so
far.  It thus leads to a mapping of the $T$ and $\mu_B$ parameters of the model
to the collision (center of mass) energy $\sqrt{s}$.  Assuming these to reflect
the true thermodynamic variables of the system and plotting these results in
the $T$-$\mu_B$ plane, one has the freeze-out curve.  We treat this curve,
solely based on precise experimental data on hadron yields, as a means to
provide the $(T, \mu)$ accessible in heavy-ion experiments as a function of the
colliding energy $\sqrt{s}$.  We use then these relevant freeze-out parameters
$T$ and $\mu_B$ in our lattice computations directly to make predictions as a
function of colliding energy.

One expects the QCD critical point to have a critical region
whose size depends on the size of the fireball as well as the critical
exponents.  The freeze-out curve may pass through it or may miss it. 
If the conditions are thus right, it may pass close enough to the 
critical point such that a study of fluctuations along it will detect its
presence.   We
define $ m_1 = T\chi^{(3)}(T,\mu_B)/\chi^{(2)}(T,\mu_B)$, $ m_3 =
{T\chi^{(4)}(T,\mu_B)}/{\chi^{(3)}(T,\mu_B)}$, and $ m_2=m_1m_3$.  Note
that the variance, skew and kurtosis of the event distribution of baryon 
number measure the various $\chi$'s above.  Furthermore,
the spatial volume cancels out in these ratios, making them
also suitable for experiments who can use their favourite proxy for it.
Usually, number of participants is preferred for that role in the 
analysis of heavy ion data.

Defining $z = \mu_B/T$, and denoting by $r_{ij}$ the estimate for radius
of convergence using $\chi_i$, $\chi_j$ as before, one has 

$$m_1 = \frac{2z}{r_{24}^2} \big[ 1 + \big(\frac{2r_{24}^2}{r_{46}^2} -1 \big) 
z^2 + \big(\frac{3r_{24}^2}{r_{46}^2 r_{68}^2} - \frac{3r_{24}^2}{r_{46}^2} +1
\big) z^4 + {\cal O}(z^6) \big]~.~$$
Similar series expressions can be written \cite{our3} down for $m_2$ and 
$m_3$.  We resum these by the Pad\`e
method described above  to construct them, since they seem to capture
the critical behaviour well in Fig. \ref{polypad}: 
$$ m_1 = z P^1_1(z^2;a,b), \qquad m_3 = \frac{1}{z} P^1_1(z^2;a',b')$$.

\begin{figure}[htb]
\includegraphics[scale=0.48]{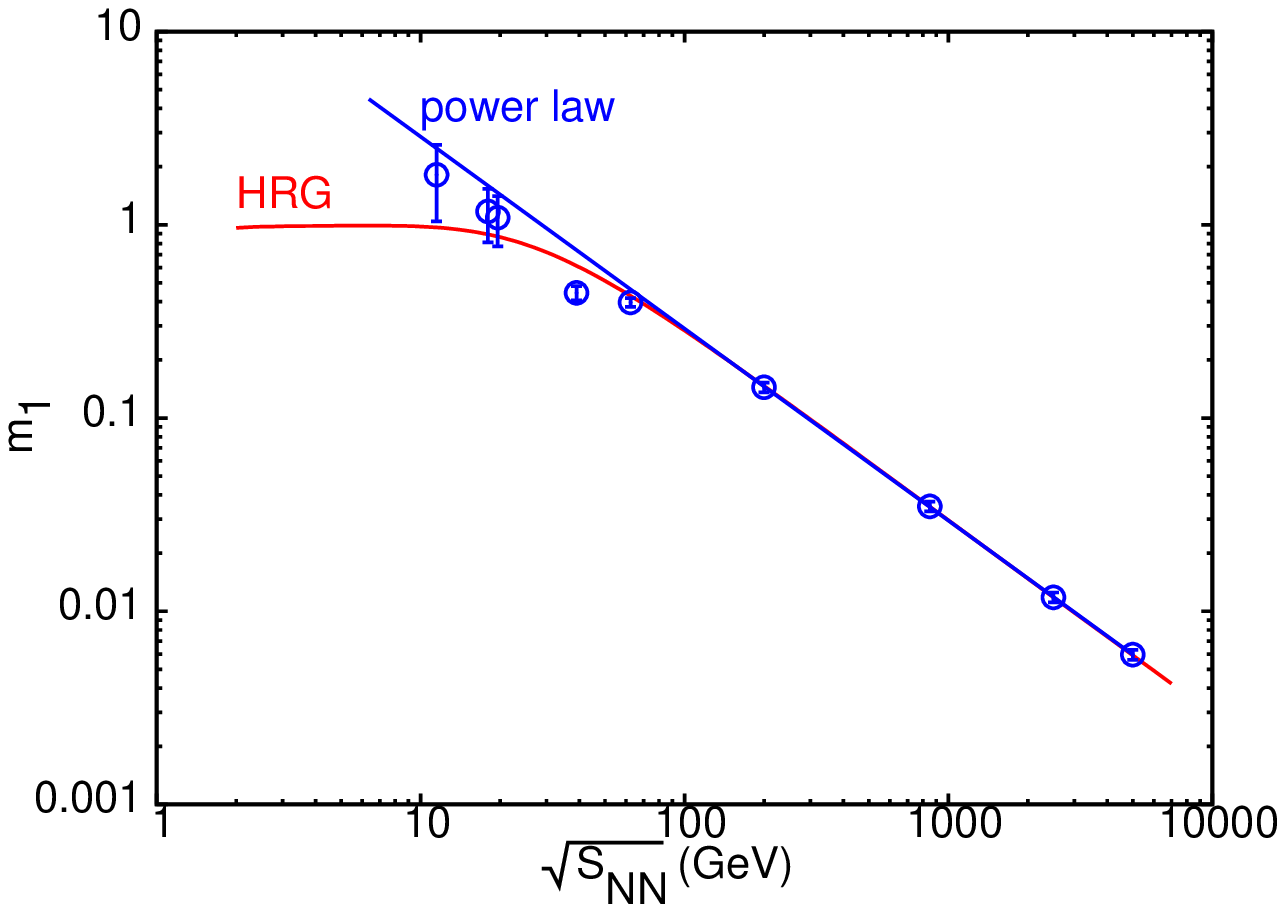} 
\includegraphics[scale=0.48]{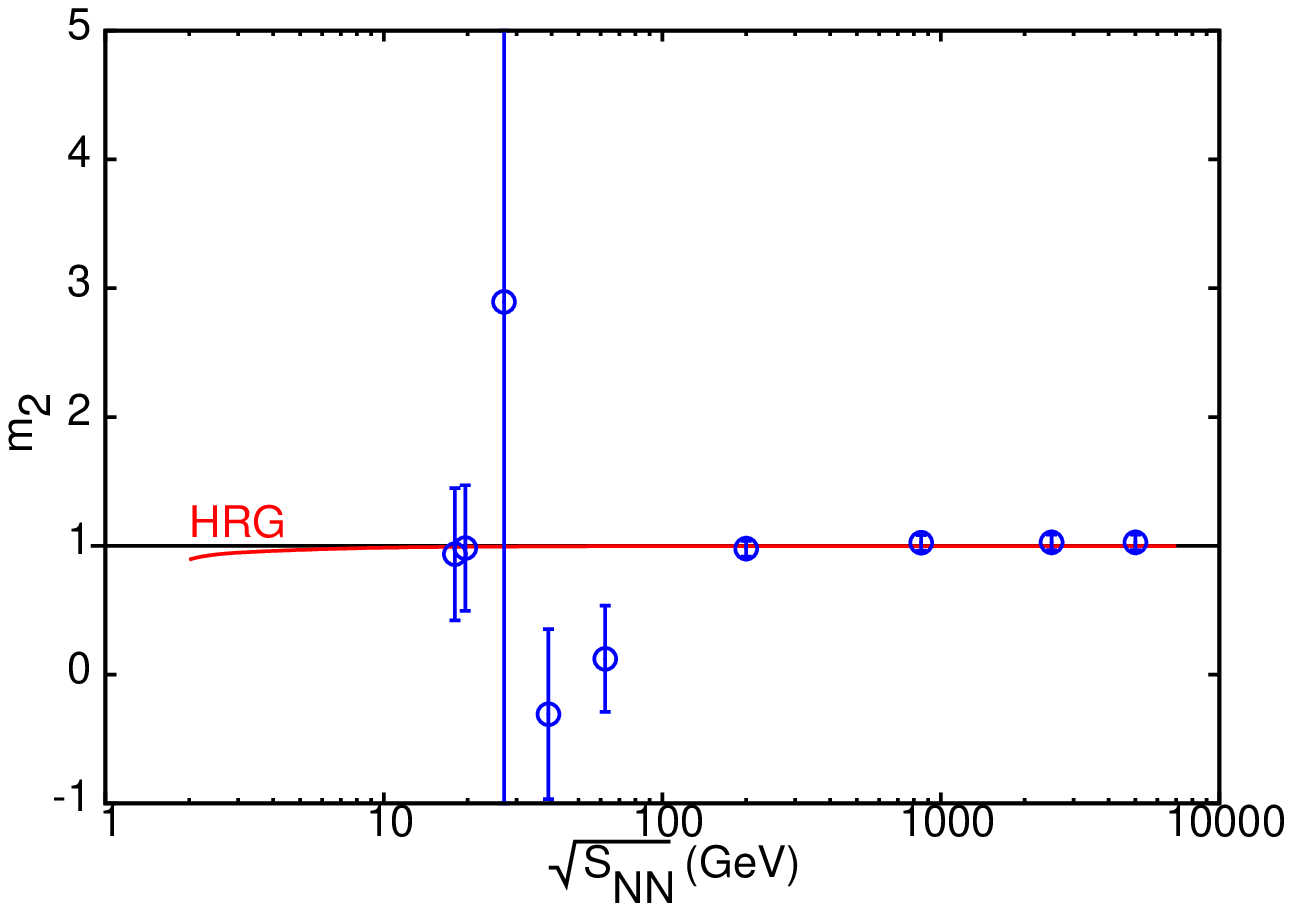} 
\caption{Results for $m_i$ on the $N_t$=6 lattice as a function of the
colliding energy. A power law fit to the high energy data and a hadron
resonance gas fit is also shown. From Ref. \cite{our3}.}
\label{m1m2m3}
\end{figure} 

Figure \ref{m1m2m3} shows our \cite{our3} results for these ratios on our finer
lattice; the results for coarser lattice are similar and can be found in Ref.
\cite{our3}.  Also shown in these figures are hadron resonance gas results 
which do not assume a critical point.  One observes a smooth and monotonic
behaviour for large $\sqrt{s}$ which is well reproduced by the hadron 
resonance gas.  Note that even in this smooth region, any
experimental comparison is exciting since it is a direct non-perturbative test
of QCD in hot and dense environment.  Earlier lattice predictions, such as
the equation of state or the transition temperature can be compared with
experiments only indirectly. Remarkably, a non-monotonic  behaviour is visible 
in Fig. \ref{m1m2m3} at our estimated critical point in all $m_i$, suggesting 
that it would be accessible to the low energy scan of RHIC at BNL. Indeed,
even if one were to be cautious in trusting the numerical precision of our
results, what should be clear that such a non-monotonic  behaviour seen at
any other nearby location would still signal the presence of QCD critical 
point.

In order to confront these results on the baryon number fluctuations with data,
one needs to address the issue of neutral baryons---neutrons---which are not
easy to detect and are thus missed.  It turns out that proton number
fluctuations suffice \cite {HaSt}.  Since the diverging correlation length at
our critical point is linked to the $\sigma$ mode which cannot mix with any
isospin modes, and the isospin susceptibility $\chi_I$ must be regular there.
Assuming protons, neutrons, pions to dominate, Ref \cite{HaSt} showed $\chi_B$
to be dominated by proton number fluctuations only.  The STAR collaboration has
recently exploited this idea and constructed the ratios $m_1$ and $m_2$ from
net proton distributions \cite{STAR}.

\begin{figure}[htb]
\hskip 1.5 cm \includegraphics[scale=0.5]{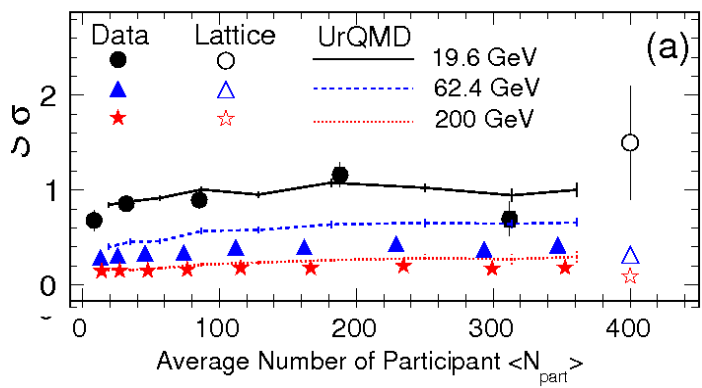}
\caption{Comparison of the STAR collaboration results with our $m_1$
results as a function of number of participants. From Ref. \cite{STAR}.}
\label{prlst}
\end{figure} 

Figure \ref{prlst} shows their results for $m_1$ against our lattice
data.  Remarkably, one observes a good agreement with our lattice results.
Hopefully, future results from the RHIC energy scan will locate the
critical point this way.  Similar agreement is also seen for $m_2$.

\section{Summary}
\label{su}

The QCD phase diagram in the $T$-$\mu_B$ plane has begun to emerge using first
principles lattice approach.  Our finer lattice results for $N_t =6$ are first
to begin the crawling towards continuum limit, suggesting a  critical point at
$\mu_B/T \sim$ 1-2.  Ratios of nonlinear susceptibilities appear to be smooth
on the freeze-out curve at large colliding energy.  Critical point leads to
structures in these $m_i$, which may be accessible in future.  STAR results on
proton number fluctuations appear to agree with our lattice QCD predictions,
making this a unique direct non-perturbative test of hot and dense QCD in
experimental tests.  So far no signs of a critical point have emerged in the
CERN and RHIC experimental results.  An interesting question is whether the
RHIC energy scan deliver it for us or whether we will need to wait for FAIR at
GSI to be operational.  At any rate, exciting future awaits these programs and
all of us interested in it.

\end{document}